# A SIMPLE ANALYSIS OF EARTHLIKE EXOPLANETS

Hristo Delev[1]

[1]Plovdiv University "Paisii Hilendarski"

**Abstract**

The astrobiology is an interdisciplinary science, combining the methods and the means of physics, biology, chemistry and astronomy. Its main purpose is to find out if the exoplanets are habitable and if so, to confirm life on them. The basic conditions for habitability are the essential ones, like these on the Earth. But additional, essential as well, exist, like the mass of the exoplanet, the atmospheric composition and its location. To find the answer we used basic molecular physics and classical mechanics knowledge. We also proposed a method for exoplanet hunting and conformation and exploration of their atmospheres. With the constantly improving techniques and apparatuses, the answer converts from "If" to "When" are we going to find extraterrestrial life, including microbial one.

**Introduction**

The biggest dream of the humankind is to find the answer to the biggest questions like "Are we alone?" and "Where do we come from?" So far, we have some hypothesizes related to the occurrence of the life on our planet. We have made experiments to confirm the spontaneous occurrence of organic molecules from non-organic ones [1]. But not on every planet could emerge life.

To answer we start from the earthly conditions like the presence of liquid water, the planetary mass and the presence of atmosphere. We know atmosphere is crucial for maintaining the water in liquid state and maintaining relatively low day-night temperature amplitude by transferring the heat. However, the atmosphere and its composition, may play the role as a cosmic shield.

The mass of the planet, on the other hand, is quite an important factor as well. By it we can judge the internal structure of the planet. On the Earth we have internal solid core and external liquid core. Due to the external molten iron core, we have magnetic field – our cosmic shield, keeping the solar wind away from surface. But by the mass we can also judge the atmospheric composition.

By the end of the past century planets orbiting other stars used to be science fiction. But in 1992 the dream came true – the first two exoplanets ever were discovered. They were reported to orbit a pulsar [2]. And few years later, in 1995, a groundbreaking discovery was reported – an exoplanet that orbits a sunlike star [3]. Since then, the number of the exoplanet has skyrocketed to more than 4 thousand and is still climbing [4].

**Mass**

According to their mass, the exoplanets are divided into several groups but these which we are going to discuss are superearths and earthsized exoplanets. The mass of the planet plays a key role for the maintenance of a thick atmosphere, rich in different gasses. We can also judge if the exoplanet has a magnetic field.

What role does the mass of the exoplanet play for the maintenance of thick atmosphere? We start from the Boltzmann distribution of the molecules by mass (1),

(1) $$n = n_0 e^{\frac{-E_p}{kT}}$$

where *Ep=mah* is the potential energy of the molecule. The formula also says that heavy gases will be located closer to the surface and their density decreases exponentially with altitude, with the temperature remaining constant. In the expression for the potential energy the surface acceleration is also included and it directly depends on the mass of the planet. In other words, the more massive the planet is, the higher potential energy the molecule has. In this case we bear in mind that the altitude and the mass of the molecule are constants. Not only does the potential energy depend on the surface acceleration, but on the mass of the molecule too. And the higher the mass of the molecule, the closer to the surface will it be. In conclusion, we can say, the higher potential energy the molecule has, the closer to the surface will it be, at a given temperature.

But in formula (1) takes part the surface temperature of the planet too. It depends on the location of the planet itself around the star or the so called Stellar habitable zone. Basically, the higher the temperature is, the more molecules with given mass will be at the given altitude.

Summarized, we can say that the higher the mass of the molecule (its potential energy), the higher the mass of the planet itself and the lower the temperature on the surface of the exoplanet, the closer to the surface the heavy gases will be. This is the main reason why oxygen is closer to the surface and by increasing the altitude its concentration decreases exponentially. Hydrogen and helium are lighter and can't remain close to the ground, while chlorine and fluorine are more massive than oxygen and get as low as possible.

The mass of the planet also plays a vital role for the existence of a liquid core. We know from the Maxwell's equations that the alternating electric field inducts alternating magnetic field and vis versa. The molten iron core plays a role like a dynamo and the flow of it inducts magnetic field.

But for a magnetic field to exist, the mass of the planet must be comparable to the mass of the Earth, for our study. Due to the high mass, the pressure in the core is enormous, so is the temperature. Take Mars for example. Its mass is 10.7% of the mass of the Earth and due to it, the planet has no magnetic field, or it is so miniscule, it can be neglected [5]. The axial rotational period also affects the magnetic field. The faster the planet rotates, the stronger magnetic field the planet has, since the flow of the liquid iron will be more intense. The magnetic field itself plays a role as a cosmic shield defending the life on our planet from the harmful solar and galactic radiation. From the latter, we can say the more massive the planet is, the closer to the surface the heavy molecules will be and the higher the chance for existence of a magnetic field is.

**Atmosphere**

The presence of atmosphere is a crucial condition for a planet to sustain life. Firstly, it is vital for the breathing and the photosynthesis, secondly it works as a heat exchanger and last but not least, the presence of a thick atmosphere keeps the water on the surface in liquid state.

By increasing the altitude, the atmospheric pressure decreases exponentially (2). And as known, the lower the atmospheric pressure is, the lower the temperature of boiling water is.

(2) $p = p_0 e^{\frac{-\mu g h}{RT}}$

At the sea level the boiling temperature of water is 371K while on mount Everest it is 339K due to the lower pressure (37.94% of the sea level pressure). At the 100km Karman limit, the theoretical limit of space, the pressure is hardly 1.67 Pa and water starts boiling even at room temperature. Solving for $h$ at $p=p_0/2$ we get about 6 km.

But not any atmosphere makes a planet habitable – its composition is a defining factor. For our study we are going to discuss the gasses like oxygen, carbon dioxide and water vapor.

The atmosphere is transparent to the visible light, infrared and radio waves and the soft UV radiation. On the other hand, it either absorbs or reflects hard UV radiation X-rays and other ionizing radiation. For example, the X-ray and hard UV rays are absorbed at 350 km altitude, while the soft UV radiation is absorbed at about 35 km altitude. When an UV quant interacts with an oxygen molecule, it dissociates and when the free oxygen atom bonds to another molecule an ozone molecule is created. The ozone in the Earth's atmosphere is crucial because it absorbs the UVb radiation which is harmful to the living organisms. And not every planet can maintain oxygen – its mass must be comparable to the Earth's. We have discussed this in the previous chapter.

Summarizing the latter, we can say the presence of oxygen is quite a good indicator the planet is habitable and protective. Stars emit in any wavelength, including UV that interact with it, and therefore we can say there is ozone, defending the planet from the UV themselves. In conclusion, we can state that the more massive the planet is, the more likely it is to find oxygen on it.

The next gas we are going to discuss is the carbon dioxide. It is responsible for the greenhouse effect on the Earth. Without it the temperature on our planet would be below the freezing temperature of the water and the surface would be frozen forever. But due to the presence of the carbon dioxide and the greenhouse effect the average temperature is about 15$^O$C [6]. Finding greenhouse gases on the exoplanets is a key indicator for greenhouse effect on them and depending on their location around the star, we can say the surface temperature is favorable for finding liquid water on it and not too high for dissolving the proteins. Of course, the concentration of carbon dioxide must be in reasonable limits, so that the greenhouse is not too intense. Nowadays we are witnesses of the global warming of the Earth due to the increasing concentration of the gas. Venus is also the best example of intense greenhouse effect since its atmosphere is 95% $CO_2$. But how do we understand the composition of the exoplanets' atmospheres?

**Transiting spectroscopy**

The best way for this is the transiting spectroscopy. Combining the transit method and the spectroscopy we can simultaneously find exoplanets, or confirm such, and with appropriate apparatuses mounted on the space telescope, we get the perfect telescope for this purpose.

The key ingredients for a greenhouse effect are carbon dioxide and IR rays. Every star emits in all the electromagnetic spectrum, including IR rays. IR rays are also a good indicator for the presence of water vapor in the atmosphere. As we know, the peak of absorption of electromagnetic radiation in water is at 1-15 μm, with its peak at 8 μm [7]. So, measuring the IR radiation, passed through the exoplanet's atmosphere, we can judge for the presence of water vapor.

This kind of spectroscopy allows us not only to find the stellar composition via the spectroscope, mounted on the space telescope but to find the spectrum of the scattered and the absorbed by the atmosphere stellar light. The passed light will give us information about the atmospheric composition. Basically, when we have a drop in the lightcurve, we can measure it and calculate the size of the exoplanet. Not only do we get the size but we measure the orbital period, respectively, the distance to the parent star and may conclude if it is in the stellar habitable zone. In addition, using the mounted spectroscope aboard the space observatory, we get a pretty good view of the stellar and the planet's atmosphere composition.

The main obstacle to this method is that it is efficient for finding big planets like gas giants and superearths. Does this mean it is excluded to find life on them? In first sight – yes. But in September 2020 a revolutionary paper was published [8]. This gives us a clue that life is possible to emerge not on rocky planets only but on gas ones too. The paradox comes from the gas giants' physics. They have no solid ground and the deeper we go, the higher the temperature gets. So, undoubtably, they are not the perfect candidates for discovering life there, but the option shouldn't be excluded. According to the study, life could emerge in their atmospheres, floating in the air and the clouds.

In addition, some of these planets have big and massive enough moons, which can be thought as individual planets. All the reflections from the latter are valid for them as well. In this case we widen our limits not only to the planets themselves but to their moons too. We have some good examples of this. Although Ganimede is a moon, it is bigger than Mercury. Enceladus and Europa are also good examples for good candidates, since they contain organic molecules in their undersurface oceans [9].

**Results and discussion**

The life is thought to be a unique occurrence, but is it. In the present study we have discussed the additional condition of the exoplanets, so that we can find live on them. The presence of atmosphere is crucial, but its composition is essential. But to have a thick enough atmosphere, the mass of the exoplanet has to be earthlike. The higher the mass, the thicker the atmosphere is and more likely it is to find heavy gases in it. Using the Boltzmann distribution, we can also see that by increasing the surface temperature the likelihood of finding heavy molecules close to the surface is also decreasing. But depending on their mass, they will be located in the lower layers of the atmosphere. Having heavy gases like oxygen helps the creation of the ozone layer and the carbon dioxide supports the greenhouse effect. Also using the combination of the transit method and the spectroscopy, we have a great tool for investigating the planetary atmosphere composition. Finding a peak of absorption of the electromagnetic waves in the interval 1-15μm is a serious clue for the presence of water and water vapor in its air. It turns out not only planets are appropriate candidates for life, but their moons too. Even if the planets are supergiants and superearths that doesn't exclude them. We now know life could emerge in their atmospheres, not mandatory on the surface. What if all the effort we put turns out to be meaningless and we do not find life on any of them? Does that mean we are all alone in the vast Universe?